\RequirePackage{fix-cm}

\documentclass[smallextended,final]{svjour3}

\usepackage{epsfig}
\usepackage{graphicx}
\usepackage{mathptmx}      
\usepackage{latexsym}
\usepackage{amssymb}
\usepackage[centertags]{amsmath}

\newcommand{\A}{\hat{a}}
\newcommand{\Ad}{\hat{a}^\dagger}
\newcommand{\B}{\hat{b}}
\newcommand{\Bd}{\hat{b}^\dagger}

\newcommand{\F}{\mathcal{F}}
\newcommand{\N}{\mathcal{N}}
\newcommand{\E}{\mathcal{E}}

\journalname{Journal of Low Temperature Physics}

\begin{document}

\title{Relaxation rates and collision integrals for Bose-Einstein condensates}
\author{Erich D. Gust       \and
        L. E. Reichl}

\institute{The Center for Complex Quantum Systems, The University of Texas at Austin, Austin, Texas 78712 \\
           Tel.: 512-471-7253 \\
           Fax: 512-471-9621 \\
           \email{egust@physics.utexas.edu \and reichl@physics.utexas.edu}
}

\date{Received: date / Accepted: date}

\maketitle

\begin{abstract}
Near equilibrium, the rate of relaxation to equilibrium  and the transport properties of excitations (bogolons) in a  dilute Bose-Einstein condensate (BEC) are determined by three collision integrals, $\mathcal{G}^{12}$, $\mathcal{G}^{22}$, and $\mathcal{G}^{31}$. All three collision integrals conserve momentum and energy during bogolon collisions, but only $ \mathcal{G}^{22}$ conserves bogolon number.  Previous works have considered the contribution of  only two collision integrals,  $ \mathcal{G}^{22}$ and $ \mathcal{G}^{12}$. In this work, we show that the third collision integral $ \mathcal{G}^{31}$ makes a significant contribution to the bogolon number relaxation rate and needs to be retained when computing relaxation properties of the BEC. We provide values of relaxation rates in a form that can be applied to a variety of dilute Bose-Einstein condensates.
\end{abstract}


%
%
\section{Introduction \label{sec:intro}}

The relaxation of a dilute gas to thermal equilibrium is governed by a kinetic equation that contains one or more collision integrals. Each collision integral represents the effect of a certain microscopic collision process on the evolution of the distribution function. Taken together, the collision integrals determine the rate of relaxation to equilibrium. In classical gases, the relaxation is governed by the Boltzmann equation and is determined by a single collision integral, ${\hat C}^{22}$, that represents an elastic binary collision and therefore explicitly conserves the number of particles, momentum and kinetic energy of the gas \cite{cerci,gust1,reichl}. Similarly, for a dilute quantum gas of bosons that is degenerate (but non-condensed), the relaxation is governed by the Uehling-Uhlenbeck equation and is determined by a modified form of ${\hat C}^{22}$ that leads to a Bose-Einstein distribution as the equilibrium state \cite{ueh,gust2}.

Analogous kinetic equations have also been derived for dilute boson gases in the presence of a Bose-Einstein condensates (BEC). In ground-breaking work, Kirkpatrick and Dorfman  \cite{dorf1}, derived a mean field kinetic equation (the KD equation) for a dilute condensed Bose gas that describes the relaxation in terms of ``collisions" between Bogoliubov excitations. A key feature of the KD equation is that the number of Bogliubov excitations (bogolons) need not be conserved during the collision process. The KD equation contains two collision integrals, which we denote as ${\cal G}^{22}$ and ${\cal G}^{12}$, that both conserve the energy and momentum of the bogolon gas. However, only ${\cal G}^{22}$, which describes an elastic binary collision between bogolons, conserves bogolon number. The collision integral ${\cal G}^{12}$ allows two (one) bogolons to enter the collision and one (two) to leave, thus accounting for the non-conservation of bogolon number during collisions. Subsequently, several authors \cite{stoof,zaremba,griffin} have studied transport properties of dilute BECs using mean field kinetic equations very similar to the KD kinetic equation with the same two collision integrals.

We have found that there is a third collision integral, ${\cal G}^{31}$, that should be included in the dilute BEC mean field kinetic equation, because it gives a dominate contribution to the relaxation of the bogolon number, and a considerable contribution to the relaxation of other important modes in the dilute BEC. We have previously derived the form of the collision integral ${\cal G}^{31}$ \cite{gust4} and found that it can be interpreted as a collision where one (three) bogolons enter and three (one) leave. The purpose of this paper is to describe and compute the effect of the collision integral ${\cal G}^{31}$ on the relaxation of dilute BECs to equilibrium.

In the all of the theories described above, the relaxation to equilibrium is caused by the internal collision processes between particles or excitations and is described by collision integrals. This is distinct from the stochastic methods \cite{cfield} which describe relaxation by contact with external reservoirs, stochastic initial conditions or phenomenological damping.

We begin in Section \ref{sec:theory}, with a summary of basic concepts underlying the analysis of the Bogolon kinetic equation. In Section \ref{sec:kineq}, we write the bogolon kinetic equation, and in Section \ref{sec:lin} we linearize the kinetic equation in terms of deviations from equilibrium. In Section \ref{sec:selfcons} we describe how the parameters in the collision integrals are obtained from a physical system. In Section \ref{sec:result}, we obtain relaxation rates and in Section \ref{sec:conc} we make some concluding remarks.

\section{Background  \label{sec:theory}}

We consider a spatially uniform system of bosons of mass $m$ that are confined to a rectangular box of volume $V$ with periodic boundary conditions. We assume that the particles interact via a contact potential $V({\bf r}_i, {\bf r}_j) = g \delta^3({\bf r}_i - {\bf r}_j)$, where ${\bf r}_i$ is the displacement of the $i^{\rm th}$ particle and $g$ is the strength of the interaction.

The Hamiltonian for this boson gas is
\begin{equation} \label{eq:H}
\hat{H} = \sum_{{\bf k}_1} \epsilon_{{\bf k}_1} \Ad_{{\bf k}_1} \A_{{\bf k}_1} + \frac{g}{2 V} \sum_{{\bf k}_1} \sum_{{\bf k}_2} \sum_{{\bf k}_3} \sum_{{\bf k}_4} \delta_{{\bf k}_1 + {\bf k}_2, {\bf k}_3 + {\bf k}_4} \Ad_{{\bf k}_1} \Ad_{{\bf k}_2} \A_{{\bf k}_3} \A_{{\bf k}_4},
\end{equation}
where $\epsilon_{{\bf k}_1} = \frac{{\hbar}^2 k_1^2}{2 m}$, $\Ad_{{\bf k}_1}$ creates a particle with momentum $\hbar {\bf k}_1$ and $\A_{{\bf k}_1}$ destroys a particle with momentum $\hbar {\bf k}_1$. The creation and annihilation operators satisfy the boson commutation relations $[\A_{{\bf k}_1}, \Ad_{{\bf k}_2}] = \delta_{{\bf k}_1, {\bf k}_2}$ where $\delta_{{\bf k}_1, {\bf k}_2}$ is the product of three Kronecker delta functions, one for each component of ${\bf k}$. The summations run over all single particle states for both positive and negative components of ${\bf k}$. To simplify notation in subsequent sections, we will let $\sum_{{\bf k}_1} \rightarrow \sum_1$, $\Ad_{{\bf k}_1} \rightarrow \Ad_1$, and $\A_{{\bf k}_1} \rightarrow \A_1$. We will keep this subscript convention for all quantities which are functions of the wavevector ${\bf k}_i$, such as $\epsilon_1 = \epsilon_{{\bf k}_1}$ and $\delta_{1,2} = \delta_{{\bf k}_1,{\bf k}_2}$.

To correctly describe a system containing a BEC, we must allow for the expectation values $\langle \Ad_{\bf k} \Ad_{-{\bf k}} \rangle$ and $\langle \A_{-{\bf k}} \A_{\bf k} \rangle$ to be non-zero at equilibrium \cite{fet}. To accomplish this, we introduce two mean fields, $\nu$ and $\Delta$, and write the unperturbed Hamiltonian $\hat{H}^0$ and interaction Hamiltonian $\hat{H}^1$ as
\begin{equation} \label{eq:H0}
\hat{H}^0 = \Xi + {\sum_i} \left[ (\epsilon_i - \mu + \nu) \Ad_i \A_i + \frac{\Delta}{2} (\Ad_i \Ad_{-i} + \A_{-i} \A_i) \right]
\end{equation}
and
\begin{equation} \label{eq:H1}
\hat{H}^1 = \frac{g}{2 V} {\sum_{ijkl}} \delta_{i+j, k+l} \Ad_i \Ad_j \A_k \A_l - {\sum_i} \left[ \nu \Ad_i \A_i + \frac{\Delta}{2} (\Ad_i \Ad_{-i} + \A_{-i} \A_i) \right] - \Xi.
\end{equation}
The mean fields $\nu$ and $\Delta$ are determined so that secular terms are eliminated from the kinetic equations \cite{frieman}. To lowest approximation, they are
\begin{equation} \label{eq:nueqAA}
\nu = \frac{2 g}{V} {\sum_i} \langle \Ad_i \A_i \rangle = \frac{2 g N}{V}
\end{equation}
and
\begin{equation} \label{eq:deeqAA}
\Delta = \frac{g}{V} {\sum_i} \langle \Ad_i \Ad_{-i} \rangle = \frac{g}{V} {\sum_i} \langle \A_{-i} \A_i \rangle.
\end{equation}
The quantity $\Xi$ is a shift in the energy of the system that is chosen so $\langle \hat{H}^1 \rangle = 0$ in equilibrium at zero temperature. Since $\Xi$ is not an operator, it will not enter into the collision integrals. As discussed in Ref. \cite{gust4}, the chemical potential, $\mu$, is set equal to $\nu - \Delta$ and the average particle density $n = \frac{N}{V}$ implicity determines the values of $\nu$ and $\Delta$.

The mean field unperturbed Hamiltonian (\ref{eq:H0}) can be diagonalized by a Bogoliubov transformation. We define the operators
\begin{equation} \label{eq:BdB}
\Bd_i = u_i \Ad_i + v_i \A_{-i} \hspace{0.5in} {\rm and} \hspace{0.5in} \B_i = u_i \A_i + v_i \Ad_{-i}
\end{equation}
with
\begin{equation}
u_i = \frac{1}{\sqrt{2}}\sqrt{\frac{\epsilon_i + \Delta}{E_i} + 1}, \hspace{1in} v_i = \frac{1}{\sqrt{2}}\sqrt{\frac{\epsilon_i + \Delta}{E_i} - 1}
\end{equation}
where
\begin{equation} \label{eq:bogolonE}
E_i = \sqrt{(\epsilon_i + \Delta)^2 - \Delta^2}.
\end{equation}
The operators $\Bd_i$ and $\B_j$ have the property that $[\B_i, \Bd_j] = \delta_{i,j}$ and in terms of them, the mean field unperturbed Hamiltonian is
\begin{equation} \label{eq:boglonH0}
\hat{H}^0 = {\sum_i} E_i \Bd_i \B_i + \frac{1}{2}{\sum_i} \left( E_i - \epsilon_i - \Delta \right) + \Xi.
\end{equation}
Therefore, $\Bd_i$ and $\B_i$ are interpreted as creation and annihilation operators for bosonic excitations, which we will refer to as bogolons. The transformation factors $u_i$ and $v_i$ are undefined at zero momentum, so we must give special treatment the operators $\Ad_0$ and $\A_0$. Following Bogoliubov, we implement the special treatment of zero momentum operators by replacing $\Ad_0$ and $\A_0$ with $\sqrt{N_0}$ where $N_0$ is the number of particles in the zero momentum state. The quantity $N_0$ is also interpreted as the number of particles in the condensate, and the condensate density can be defined as $n_0 = \frac{N_0}{V}$.

\section{Collision Integrals for a Condensed Bose Gas \label{sec:kineq}}

Though we consider a dilute BEC that is spatially uniform, the momentum (energy) distribution of the bogolons may be out of equilibrium. The kinetic equation governing the relaxation of the bogolon momentum distribution to equilibrium can be derived from the Hamiltonians presented in the preceding section. We have derived the mean field kinetic equation for dilute BECs using the approach of Peletminksii and Yatsenko (PY) \cite{pel1,pel2}. The derivation of the collision integrals can be found in Ref. \cite{gust4}. We obtain a kinetic equation for the relaxation of bogolon momentum distribution that is nearly identical to the KD equation, but contains the a third collision integrals, ${\cal G}^{31}$, in addition to the known collision integrals ${\cal G}^{22}$ and ${\cal G}^{12}$. Taking the thermodynamic limit by letting $V \to \infty$ and $N \to \infty$ while keeping $n = \frac{N}{V}$ constant, the kinetic equation for an infinite, uniform bogolon gas can be written
\begin{equation} \label{eq:bogokineq}
\frac{d \N({\bf k})}{d t} = \mathcal{G}^{12}_{\bf k} \{\N\} + \mathcal{G}^{22}_{\bf k} \{\N\} + \mathcal{G}^{31}_{\bf k} \{\N\},
\end{equation}
where $\N({\bf k}) \equiv \langle \Bd_{\bf k} \B_{\bf k} \rangle $ is the expected number of bogolons with momentum ${\bf k}$. The bogolon collision integrals are given by (we include ${\cal G}^{22}$ and ${\cal G}^{12}$ for completeness and comparison)
\begin{equation} \label{eq:G12}
\begin{split}
\mathcal{G}^{12}_{{\bf k}_1} \{\N\} =& \frac{4 \pi N_0 g^2}{(2\pi)^3 \hbar V} \int d {\bf k}_2 d {\bf k}_3  \\ \times
\Big[ 2 \delta^3 &({\bf k}_1 + {\bf k}_2 - {\bf k}_3) \delta(E_1 + E_2 - E_3) (W^{12}_{1,2,3})^2 (\F_1 \F_2 \N_3 - \N_1 \N_2 \F_3) \\
+ \delta^3 &({\bf k}_1 - {\bf k}_2 - {\bf k}_3) \delta(E_1 - E_2 - E_3) (W^{12}_{3,2,1})^2 (\F_1 \N_2 \N_3 - \N_1 \F_2 \F_3) \Big],
\end{split}
\end{equation}
\begin{equation} \label{eq:G22}
\begin{split}
\mathcal{G}^{22}_{{\bf k}_1} \{\N\} =& \frac{4 \pi g^2}{(2\pi)^6 \hbar} \int d{\bf k}_2 d{\bf k}_3 d{\bf k}_4 \\  \times
 &\delta^3 ({\bf k}_1 + {\bf k}_2  - {\bf k}_3 - {\bf k}_4) \delta(E_1 + E_2 - E_3 - E_4)  \\ \times
&(W^{22}_{1,2,3,4})^2 (\F_1 \F_2 \N_3 \N_4 - \N_1 \N_2 \F_3 \F_4)
\end{split}
\end{equation}
and
\begin{equation} \label{eq:G31}
\begin{split}
\mathcal{G}^{31}_{{\bf k}_1} \{\N\} = \frac{4 \pi g^2}{(2\pi)^6 \hbar} &\int d{\bf k}_2 d{\bf k}_3 d{\bf k}_4  \\ \times
\Big[ \frac{1}{3} \delta^3 &({\bf k}_1 - {\bf k}_2 - {\bf k}_3 - {\bf k}_4) \delta(E_1 - E_2 - E_3 - E_4)  \\ \times
 &(W^{31}_{1,2,3,4})^2 (\F_1 \N_2 \N_3 \N_4 - \N_1 \F_2 \F_3 \F_4) \\
 + \delta^3 &({\bf k}_1 + {\bf k}_2 + {\bf k}_3 - {\bf k}_4) \delta(E_1 + E_2 + E_3 - E_4)  \\ \times
 &(W^{31}_{4,3,2,1})^2 (\F_1 \F_2 \F_3 \N_4 - \N_1 \N_2 \N_3 \F_4) \Big],
\end{split}
\end{equation}
where $\F_i = 1 + \N_i$.

The weighting functions are given in terms of $u_i$ and $v_i$ by
\begin{equation} \label{eq:W12}
W^{12}_{1,2,3} = u_1 u_2 u_3 - u_1 v_2 u_3 - v_1 u_2 u_3 + u_1 v_2 v_3 + v_1 u_2 v_3 - v_1 v_2 v_3,
\end{equation}
\begin{equation} \label{eq:W22}
W^{22}_{1,2,3,4} = u_1 u_2 u_3 u_4 + u_1 v_2 u_3 v_4 + u_1 v_2 v_3 u_4 + v_1 u_2 u_3 v_4 + v_1 u_2 v_3 u_4 + v_1 v_2 v_3 v_4
\end{equation}
and
\begin{equation} \label{eq:W31}
W^{31}_{1,2,3,4} = u_1 u_2 u_3 v_4 + u_1 u_2 v_3 u_4 + u_1 v_2 u_3 u_4 + v_1 v_2 v_3 u_4 + v_1 v_2 u_3 v_4 + v_1 u_2 v_3 v_4.
\end{equation}
Each of these weighting functions has specific symmetry with respect to interchanges of its indices that is shared by its corresponding collision integral. The collision integrals ${\cal G}^{12}$ and ${\cal G}^{22}$ are identical those given in Ref. \cite{dorf1} and discussed extensively in Ref. \cite{griffin}. It is not difficult to show that a Bose-Einstein distribution in the bogolon energies,
\begin{equation}
\N^0_i = \frac{1}{e^{\frac{E_i}{k_B T}} - 1},
\end{equation}
is the long-time steady-state solution of Eq. (\ref{eq:bogokineq}). In Ref. \cite{gust4} we prove that these collision integrals conserve average particle number even through they do not conserve bogolon number.

\section{Linearized Collision Operators \label{sec:lin}}

The relaxation rates of the bogolon gas are defined as the rates of pure exponential decay for infinitesimal perturbations about equilibrium. To determine the perturbations that exhibit pure exponential decay and their associated rates, we linearize the collision integrals about equilibrium to obtain the linearized collision operators. Replacing the collision integrals in the kinetic equation with the linearized collision operators results in a linear kinetic equation that can be cast as an eigenvalue equation. The eigenvalues of this equation give the relaxation rates and the eigenfunctions give the characteristic relaxation modes of the perturbations.

We linearize the collision integrals by writing the bogolon distribution as
\begin{equation}
\N_i(t) = \N_i^0 + \N_i^0 \F_i^0 \phi_i(t)
\end{equation}
where $\phi_i(t)$ represents deviations from absolute equilibrium and decays to zero as $t \to \infty$. We assume that the gas is very close to equilibrium and neglect terms beyond first order in $\phi_i(t)$. The factor of $\N_i^0 \F_i^0$ multiplying $\phi_i(t)$ simplifies the eventual form of the linearized collision operators. We also introduce a dimensionless momentum vector ${\bf c} = \frac{\hbar}{\sqrt{2 m k_B T}} {\bf k}$. In terms of ${\bf c}$ the integrals become dimensionless, and the physical quantities can be combined into an overall rate constant with units of inverse time. Using $g = \frac{4\pi\hbar^2 a}{m}$ where $a$ is the measured s-wave scattering length, this rate constant takes the form
\begin{equation} \label{eq:gamma}
\gamma = \frac{8 m a^2 (k_B T)^2}{\pi \hbar^3}.
\end{equation}
To parameterize the collision integrals in terms of dimensionless quantities, we define $b = \frac{\Delta}{k_B T}$ as the dimensionless ratio between the mean field and the temperature. Since ${\cal G}^{12}$ contains $N_0$ in its coefficient, we introduce a third dimensionless quantity,
\begin{equation} \label{eq:alpha}
\alpha = n_0 \left(\frac{2 \pi^2 \hbar^2}{m k_B T}\right)^{3/2},
\end{equation}
so that the overall coefficient of ${\cal G}^{12}$ is $\alpha \gamma$.

After the collision integrals have been linearized, we can rename the integration variables and use the symmetries of the weighting functions (\ref{eq:W12}, \ref{eq:W22}, \ref{eq:W31}) to write the kinetic equation as
\begin{equation} \label{eq:keq}
\begin{split}
\frac{d \phi({\bf c}_1, t)}{d t} = - &\gamma M({\bf c}_1) \phi({\bf c}_1, t) \\
- &\frac{\gamma}{\pi^2 \F^0_1} \int d{\bf c}_2 \N^0_2 \phi({\bf c}_2, t) \Big[ 2 \alpha T_A({\bf c}_1, {\bf c}_2) - 2 \alpha T_B({\bf c}_1, {\bf c}_2) - 2 \alpha T_B({\bf c}_2, {\bf c}_1) \\
+ &Q_A({\bf c}_1, {\bf c}_2) - 2 R_A({\bf c}_1, {\bf c}_2) + 2 Q_B({\bf c}_1, {\bf c}_2) - Q_C({\bf c}_1, {\bf c}_2) - Q_C({\bf c}_2, {\bf c}_1) \Big]
\end{split}
\end{equation}
where
\begin{equation} \label{eq:dynrates}
\begin{split}
M({\bf c}_1) = \frac{1}{\pi^2 \F^0_1} \int d{\bf c}_2 \N^0_2 \Big[
 &2 \alpha T_A({\bf c}_1, {\bf c}_2) + \alpha T_B({\bf c}_1, {\bf c}_2) + Q_A({\bf c}_1, {\bf c}_2) \\
&+ Q_B({\bf c}_1, {\bf c}_2) + \frac{1}{3} Q_C({\bf c}_1, {\bf c}_2) \Big]
\end{split}
\end{equation}
and
\begin{equation} \label{eq:TA}
T_A({\bf c}_1, {\bf c}_2) = \int d{\bf c}_3 \delta^3({\bf c}_1 + {\bf c}_2 - {\bf c}_3) \delta(\E_1 + \E_2 - \E_3) (W^{12}_{1,2,3})^2 \F^0_3,
\end{equation}
\begin{equation} \label{eq:TB}
T_B({\bf c}_1, {\bf c}_2) = \frac{\F^0_2}{\N^0_2} \int d{\bf c}_3 \delta^3({\bf c}_1 - {\bf c}_2 - {\bf c}_3) \delta(\E_1 - \E_2 - \E_3) (W^{12}_{3,2,1})^2 \F^0_3,
\end{equation}
\begin{equation} \label{eq:QA}
\begin{split}
Q_A({\bf c}_1, {\bf c}_2) =& \int d{\bf c}_3 d{\bf c}_4 \delta^3({\bf c}_1 + {\bf c}_2 - {\bf c}_3 - {\bf c}_4)  \\ \times
& \delta(\E_1 + \E_2 - \E_3 - \E_4)  (W^{22}_{1,2,3,4})^2  \F^0_3 \F^0_4,
\end{split}
\end{equation}
\begin{equation} \label{eq:RA}
\begin{split}
R_A({\bf c}_1, {\bf c}_2) =& \frac{\F^0_2}{\N^0_2} \int d{\bf c}_3 d{\bf c}_4 \delta^3({\bf c}_1 - {\bf c}_2 + {\bf c}_3 - {\bf c}_4)
 \\ \times  & \delta(\E_1 - \E_2 + \E_3 - \E_4) (W^{22}_{1,3,2,4})^2 \N^0_3 \F^0_4,
\end{split}
\end{equation}
\begin{equation} \label{eq:QB}
\begin{split}
Q_B({\bf c}_1, {\bf c}_2) =& \int d{\bf c}_3 d{\bf c}_4 \delta^3({\bf c}_1 + {\bf c}_2 + {\bf c}_3 - {\bf c}_4) \\ \times
& \delta(\E_1 + \E_2 + \E_3 - \E_4) (W^{31}_{4,3,2,1})^2 \N^0_3 \F^0_4,
\end{split}
\end{equation}
\begin{equation} \label{eq:QC}
\begin{split}
Q_C({\bf c}_1, {\bf c}_2) =& \frac{\F^0_2}{\N^0_2} \int d{\bf c}_3 d{\bf c}_4 \delta^3({\bf c}_1 - {\bf c}_2 - {\bf c}_3 - {\bf c}_4)
 \\ \times & \delta(\E_1 - \E_2 - \E_3 - \E_4) (W^{31}_{1,2,3,4})^2 \F^0_3 \F^0_4.
\end{split}
\end{equation}
In these expressions, $\E_i = \sqrt{c_i^4 + 2 b c_i^2}$.

The kinetic equation (\ref{eq:keq}) can now be cast as a general integral equation of the form
\begin{equation} \label{eq:keq3d}
- \frac{d \phi({\bf c}_1, t)}{d (\gamma t)} = M({\bf c}_1) \phi({\bf c}_1, t)
+ \int d{\bf c}_2 \frac{\N^0_2}{\F^0_1} K({\bf c}_1, {\bf c}_2) \phi({\bf c}_2, t).
\end{equation}
This equation can be simplified by expanding the angular dependence of the function $\phi({\bf c}, t)$ in spherical harmonics. To also make the resulting kernel symmetric under interchange of its arguments, we expand $\phi({\bf c}, t)$ as
\begin{equation}
\phi({\bf c}, t) = \frac{1}{c \sqrt{\N^0(c) \F^0(c)}} \sum\limits_{l = 0}^\infty \sum_{m = -l}^{l} \psi_{l,m}(c, t) Y_{l}^{m}(\hat{\bf c}).
\end{equation}
Here, $Y_{l}^{m}(\hat{\bf c})$ is a spherical harmonic and $\psi_{l,m}(c, t)$ is an expansion coefficient that is a function of $c$. Substitution of this expansion into Eq. (\ref{eq:keq3d}) leads to
\begin{equation} \label{eq:keqlm}
\begin{split}
- \frac{d \psi_{l,m}(c_1, t)}{ d (\gamma t) } = M(c_1) \psi_{l,m}(c_1, t) +& \int\limits_0^\infty d c_2 \sum_{l',m'}
c_1 c_2 \sqrt{\frac{\N^0(c_1) \N^0(c_2)}{\F^0(c_1) \F^0(c_2)}} \psi_{l',m'}(c_2, t) \\
&\times \int d\Omega_1 d\Omega_2 Y_l^{*m}(\hat{\bf c}_1) K({\bf c}_1, {\bf c}_2) Y_{l'}^{m'} (\hat{\bf c}_2).
\end{split}
\end{equation}
Since $K({\bf c}_1, {\bf c}_2)$ is continuous and rotationally invariant, it is a continuous function of $c_1$, $c_2$ and $\hat{\bf c}_1 \cdot \hat{\bf c}_2$ only, and therefore can be represented by a Legendre series of the form
\begin{equation}
K({\bf c}_1, {\bf c}_2) = \sum_l \tilde{K}_l(c_1, c_2) P_l(\hat{\bf c}_1 \cdot \hat{\bf c}_2).
\end{equation}
Using the spherical harmonic addition theorem, we can deduce that
\begin{equation}
\int d\Omega_1 d\Omega_2 Y_l^{*m}(\hat{\bf c}_1) K({\bf c}_1, {\bf c}_2) Y_{l'}^{m'} (\hat{\bf c}_2) =
2 \pi \delta_{l,l'} \delta_{m,m'} \int\limits_{-1}^1 d(\hat{\bf c}_1 \cdot \hat{\bf c}_2) P_l(\hat{\bf c}_1 \cdot \hat{\bf c}_2) K({\bf c}_1, {\bf c}_2)
\end{equation}
and therefore Eq. (\ref{eq:keqlm}) reduces to
\begin{equation} \label{eq:eig}
- \frac{d \psi_{l,m}(c_1, t)}{d (\gamma t)} = M(c_1) \psi_{l,m}(c_1, t) + \int\limits_0^\infty d c_2 K_l(c_1, c_2) \psi_{l,m}(c_2, t)
\end{equation}
where
\begin{equation} \label{eq:Kl}
K_l(c_1, c_2) = 2 \pi c_1 c_2 \sqrt{\frac{\N^0(c_1) \N^0(c_2)}{\F^0(c_1) \F^0(c_2)}} \int\limits_{-1}^1 d(\hat{\bf c}_1 \cdot \hat{\bf c}_2) P_l(\hat{\bf c}_1 \cdot \hat{\bf c}_2) K({\bf c}_1, {\bf c}_2).
\end{equation}
We have reduced the three-dimensional integral equation (\ref{eq:keq3d}) to an uncoupled set of one-dimensional integral equations (\ref{eq:eig}) indexed by $l$ and $m$.

Now let us define an eigenvalue $\lambda^n_{l,m}$ and an eigenfunction $\psi^n_{l,m}(c)$ as the pair which satisfies
\begin{equation} \label{eq:eigdef}
\lambda^n_{l,m} \psi^n_{l,m}(c_1) = M(c_1) \psi^n_{l,m}(c_1) + \int\limits_0^\infty d c_2 K_l(c_1, c_2) \psi^n_{l,m}(c_2).
\end{equation}
These eigenfunctions have the orthogonality relation
\begin{equation}
\int\limits_0^\infty d c \psi^n_{l,m}(c) \psi^{n'}_{l,m}(c) = \delta_{n,n'}.
\end{equation}
Note that the kernel $K_l(c_1, c_2)$ does not depend on $m$, so the eigenvalues and eigenfunctions are also independent of $m$. We shall drop the $m$ dependence of the eigenvalues $\lambda^n_{l,m}$ and eigenfunctions $\psi^n_{l,m}(c)$ where it will be understood that these quantities are $2 l + 1$ - fold degenerate.

Using the property (\ref{eq:eigdef}) of the eigenfunctions $\psi^n_l(c)$, Eq. (\ref{eq:eig}) can be solved and the evolution of the deviation function $\phi({\bf c}, t)$ can be written in terms of the eigenvalues and eigenfunctions as
\begin{equation}
\phi({\bf c}, t) = \frac{1}{c \sqrt{\N^0(c) \F^0(c)}} \sum\limits_{n = 0}^\infty \sum\limits_{l = 0}^\infty \sum_{m = -l}^{l} A^n_{l,m} e^{-\lambda^n_l \gamma t} \psi^n_l(c) Y_{l}^{m}(\hat{\bf c}),
\end{equation}
where the expansion coefficients $A^n_{l,m}$ depend on the initial deviation from equilibrium as described by $\phi({\bf c}, t = 0)$.

The eigenvalue equations for both the Boltzmann equation \cite{gust1} and the Uehling-Uhlenbeck equation \cite{gust2} have the same structure as Eq. (\ref{eq:eig}) and consequently they have a maximum discrete eigenvalue to which all other discrete eigenvalues converge. The existence of this maximum discrete eigenvalue for the Boltzmann equation has been proven \cite{kus} and its value was shown to be equal to the minimum value of the function $M(c)$. For the BEC, we denote this maximum discrete eigenvalue as $\lambda_M$.

The fact that energy is conserved implies that there is an $l = 0$ eigenfunction proportional to $\E(c)$ with zero eigenvalue. Similarly, the fact that momentum is conserved implies that there is an $l = 1$ eigenfunction proportional to $c$ with zero eigenvalue. This allows us to deduce that
\begin{equation}
\psi^{n=0}_{l=0}(c) \propto c \E(c) \sqrt{\N^0(c) \F^0(c)}
\end{equation}
and
\begin{equation}
\psi^{n=0}_{l=1}(c) \propto c^2 \sqrt{\N^0(c) \F^0(c)}.
\end{equation}
The normalization constant of these eigenfunctions must be determined numerically.

We can use the eigenfunctions $\psi^n_l(c)$ to write an expansion for the bogolon distribution in the linear regime as
\begin{equation} \label{eq:modeexp}
\N({\bf c}) = \N^0({\bf c}) + \frac{1}{c^2 \E(c)} \sum_{n = 0}^\infty \sum_{l = 0}^\infty \sum_{m = -l}^l \mathcal{A}^n_{l,m} e^{-\lambda^n_l \gamma t} \psi^0_0 (c) \psi^n_l(c) Y_l^m(\hat{\bf c}),
\end{equation}
where
\begin{equation}
\mathcal{A}^n_{l,m} = \int d {\bf c} \E(c) \left[\N({\bf c}) - \N^0(c)\right] \frac{\psi^n_l(c)}{\psi^0_0(c)} Y_l^{*m}(\hat{\bf c})
\end{equation}
The expansion coefficients $\mathcal{A}^0_{,0,0}$ and $\mathcal{A}^0_{1,m}$ are exactly zero because the modes $\psi^0_0(c)$ and $\psi^0_1(c)$ are the modes associated with energy and momentum conservation and do not decay. Their contributions are already contained in the equilibrium distribution $\N^0({\bf c})$, which is a stationary state of the kinetic equation.

\section{Equilibrium Properties of the Condensed Bose Gas  \label{sec:selfcons}}

Because the collision operators are linearized about equilibrium, the eigenvalues of the linearized collision operators will depend on detailed properties of the equilibrium state. Experimentally, the equilibrium state is specified by the particle properties $m$ and $a$, the total density $n$, and the temperature $T$. To connect our calculation to a physical system, we must be able to relate the parameters $b$ and $\alpha$ to these physical properties. We do this by using equations (\ref{eq:nueqAA}) and (\ref{eq:deeqAA}) to calculate the condensate density $n_0$ and the mean field strength $\Delta$. The dimensionless parameters $b$ and $\alpha$, which determine the dimensionless eigenvalues, can then be calculated from $\Delta$ and $n_0$.

For a low density gas, the simplest consistent approximation to equations (\ref{eq:nueqAA}) and (\ref{eq:deeqAA}) is the ``Popov approximation" \cite{griffin,popov}. This approximation has been shown \cite{dodd,hutch} to reproduce the experimentally measured properties of the dilute BECs very well for temperatures below about 60\% of $T_C$, the critical temperature for Bose-Einstein condensation. For the low densities and temperatures that we consider, the mean field values in the Popov approximation differ from more detailed approximations \cite{burnett,morgan} by only a few percent. In what follows, we use the Popov approximation to determine the equilibrium properties of the Bose-condensed gas.

Two dimensionless quantities which are readily available for this system are the reduced temperature $\tau = T / T_C$ and the ``gas parameter" $\eta = n a^3$. These quantities are also quite convenient for calculation, since they allow equation (\ref{eq:nueqAA}) and (\ref{eq:deeqAA}) to be written in terms of only dimensionless quantities. In the thermodynamic limit discussed in Sec. \ref{sec:kineq} and using the Popov approximation, Eq. (\ref{eq:nueqAA}) can be written as
\begin{equation} \label{eq:scdim}
\eta = \eta_0
+ \sqrt{8} \eta_0^{3/2} F \left( \left( \frac{\zeta}{\eta} \right)^{2/3} \frac{2 \eta_0}{\tau} \right)
+ \frac{8}{3 \sqrt{\pi}} \eta_0^{3/2}
\end{equation}
and Eq. (\ref{eq:deeqAA}) can be written as
\begin{equation} \label{eq:b2}
b = \left( \frac{\zeta}{\eta} \right)^{2/3} \frac{2 \eta_0}{\tau},
\end{equation}
where $\eta_0 = n_0 a^3$, $\zeta = \zeta\left( \frac{3}{2} \right) \approx 2.612$ and
\begin{equation}
F(b) = \frac{4}{\sqrt{\pi}} \int\limits_0^\infty d x \frac{x^2 \left(x^2 + 1\right)}{\sqrt{x^4 + 2 x^2} \left( e^{b \sqrt{x^4 + 2 x^2}} - 1 \right) }.
\end{equation}

Once Eq. (\ref{eq:scdim}) has been solved numerically for $\eta_0$ with a given value of $\eta$ and $\tau$, the parameters entering the collision integrals can be calculated from Eq. (\ref{eq:b2}) and the expressions
\begin{equation} \label{eq:alpha2}
\alpha = \pi^{3/2} \zeta \frac{\eta_0}{\eta} \tau^{-3/2}
\end{equation}
and
\begin{equation} \label{eq:gamma2}
\gamma = \frac{32 \pi \hbar}{m a^2} \eta^{4/3} \tau^2
\end{equation}
which are obtained from Eqs. (\ref{eq:gamma}), (\ref{eq:alpha}) and the definition of $T_C$.

These equations show that if two systems have the same values of $n a^3$ and $T / T_C$, their condensate fractions $(n_0 / n)$ and their dimensionless eigenvalues will also be the same. However, since $\gamma$ includes $m$ and $a$ explicitly, the relaxation rates will differ. In fact, Eq. (\ref{eq:gamma2}) implies an important scaling behavior: If two different systems (called $A$ and $B$) have the same values of $n a^3$ and $T / T_C$, then $R_A m_A a_A^2 = R_B m_B a_B^2$ where $R$ is any characteristic relaxation rate. This enables us to predict, for example, that the relaxation rates of $^{87}{\rm Rb} ~ (a = 105 ~ a_0)$ at $T = 7.26 ~ n {\rm K}$ and $1.44 \times 10^{19} ~ {\rm m}^{-3}$ will be approximately $7.26$\% of the relaxation rates of $^{23}{\rm Na} ~ (a = 55 ~ a_0)$ at $T = 0.1 ~ \mu {\rm K}$ and $n = 1.0 \times 10^{20} ~ \rm{m}^{-3}$. In this manner, we can use results at a specific value of $n a^3$ and $T / T_C$ to make predictions for any system with the same values of $n a^3$ and $T / T_C$ as long as its mass and scattering length are known.

\begin{figure*}[h]
\includegraphics[width=0.9\textwidth]{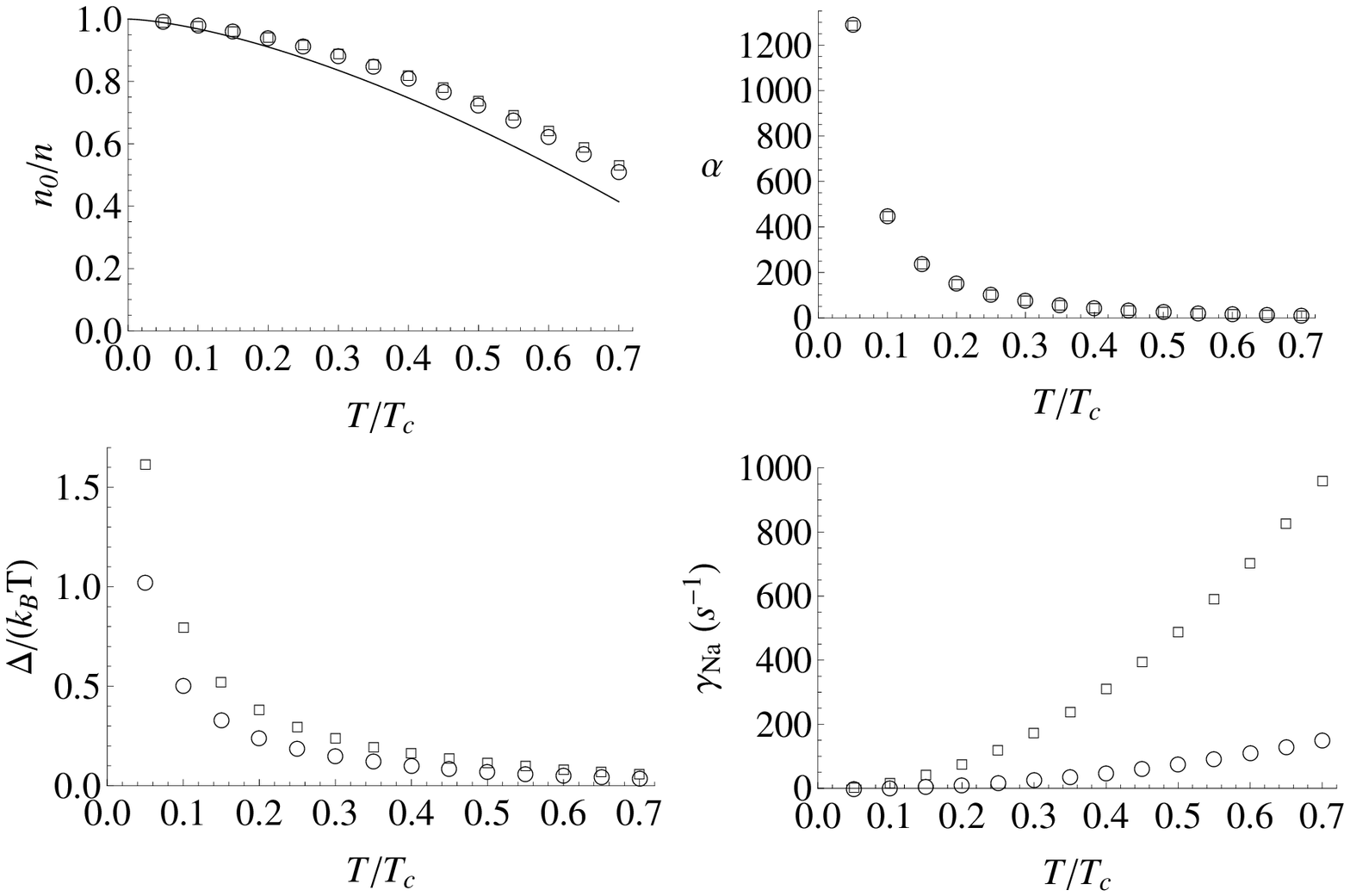}
\caption{Variation of the condensate fraction, dimensionless parameters $b$ and $\alpha$ and the overall rate coefficient $\gamma$ for $^{23}{\rm Na}$ as the temperature is increased. All quantities are computed in the Popov approximation. Circles represent a gas parameter of $n a^3 = 2.5 \times 10^{-6}$ and squares represent $n a^3 = 1.0 \times 10^{-5}$. The solid line is the condensate fraction of an ideal Bose gas.} \label{fig1}
\end{figure*}

To ensure that our choice of the parameters $n a^3$ and $T / T_C$ coincides with experimentally accessible systems, we use a dilute atomic gas of $^{23}{\rm Na}$ \cite{cornell} in the lower hyperfine state with $a = 55 ~ a_0$ \cite{hei} where $a_0$ is the Bohr radius. The calculations are carried out at a density corresponding to a gas parameter of $n a^3 = 2.5 \times 10^{-6}$. Figure \ref{fig1} shows the variation of the values of $n_0 / n$, $b$, $\alpha$ and $\gamma$ with temperature as computed using the Popov approximation. For comparison, we have also plotted points corresponding to $n a^3 = 1.0 \times 10^{-5}$.

\section{Calculation of Relaxation Rates \label{sec:result}}

In this section, we explain our numerical method for obtaining eigenvalues and eigenvectors from Eq. (\ref{eq:eig}) and give the results of the calculation. To discretize Eq. (\ref{eq:eig}) so that it can be solved numerically, we use a finite ordinate method \cite{shiz}. In this method, a quadrature scheme is chosen for evaluating integrals on the interval $0 ~ {\leq} ~ c ~ {\leq} ~ \infty$ and this scheme is applied directly to the integral in Eq. (\ref{eq:eig}). If we then demand that the eigenvalue equation is satisfied when evaluated at each quadrature point, we obtain the matrix equation
\begin{equation} \label{eq:disceig}
\lambda^n_l \psi^n_l(x_i) = M(x_i) \psi^n_l(x_i) + \sum_{j = 1}^{N_Q} w_j K_l(x_i, x_j) \psi^n_l(x_j)
\end{equation}
where $N_Q$ is the total number of points in this quadrature and $x_i$ and $w_i$ are the quadrature points and weights, respectively. This discretization method has the advantage of speed and simplicity and is quite reliable, particularly if an appropriate quadrature scheme is chosen. For our calculations, we chose to split the integration region into $0 ~ \leq ~ c_2 ~ \leq ~ 2$ and $2 ~ < ~ c_2 ~ \leq ~ 12$ and use a composite seven-point gaussian quadrature method in each region. This quadrature scheme captures the small-scale and large-scale features of the kernel.

In order to gauge the accuracy of the eigenvalues we perform the calculation several times, each with a larger total number of quadrature points $N_Q$. The eigenvalues that we report are found by performing a regression of the eigenvalues versus $1/N_Q$ and extrapolating to $1/N_Q = 0$. This gives us an estimate of the truncation error in the eigenvalues. For all non-zero eigenvalues, this truncation error is less than 1\% of the eigenvalue. For the zero-eigenvalues, we found that a 95\% confidence interval based on this truncation error always includes zero. We must note that the truncation error only estimates the error of using a finite size matrix, and does not address the error involved in the computation of matrix elements.

To calculate the matrix elements $K_l(x_i, x_j)$, we use a Gauss-Kronrod (G7-K15) global adaptive quadrature scheme with a relative tolerance of $10^{-7}$ \cite{kah}. Appendix \ref{ax:colop} contains an in-depth explanation of how we evaluate $K_l(x_i, x_j)$ given its definition in Eq. (\ref{eq:Kl}).

\begin{figure*}[h]
\includegraphics[width=0.5\textwidth]{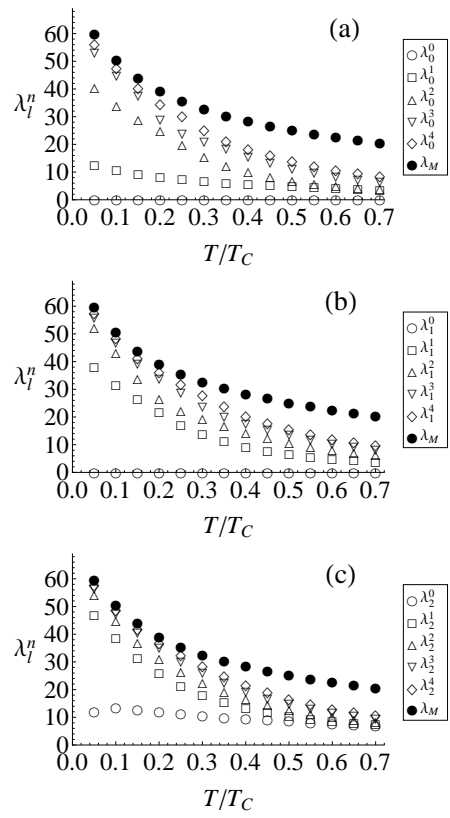}
\caption{Dimensionless eigenvalues $\lambda_l^n$ as a function of reduced temperature $T / T_C$ for $l = 0$ (a), $l = 1$ (b) and $l = 2$ (c). Solid circles indicate the maximum discrete eigenvalue $\lambda_M$. All quantities are computed in the Popov approximation at $n a^3 = 2.5 \times 10^{-6}$.} \label{fig2}
\end{figure*}

In Fig. \ref{fig2} we plot several of the dimensionless eigenvalues obtained at varying temperature. These eigenvalues are computed for $n a^3 = 2.5 \times 10^{-6}$. Note the presence of a single zero eigenvalues in the $l = 0$ plot, showing energy conservation, and the presence of another single zero eigenvalues in the $l = 1$ plot, showing momentum conservation. Analysis of the numerically calculated eigenfunctions associated with these zero eigenvalues supports this interpretation. The absence of any other zero eigenvalues confirms the fact that bogolon number is not conserved. We also find that the dimensionless eigenvalues appear to converge to $\lambda_M$ from below.

\begin{figure*}[h]
\includegraphics[width=0.9\textwidth]{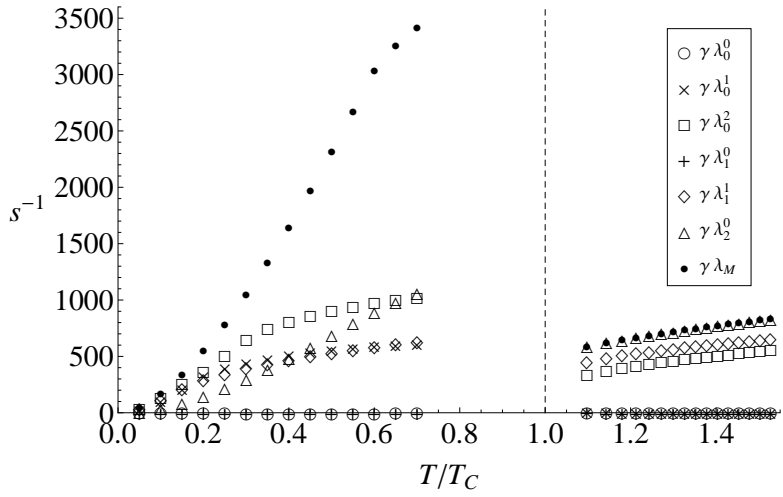}
\caption{Relaxation rates $\gamma \lambda_l^n$ versus temperature for several of the lowest eigenvalues. All quantities are computed in the Popov approximation for $^{23}{\rm Na}$ at $n a^3 = 2.5 \times 10^{-6}$. For comparison, relaxation rates of the Uehling-Uhlenbeck equation are plotted for $T > T_C$.} \label{fig3}
\end{figure*}

Figure \ref{fig3} shows the physical relaxation rates of the some of the lower relaxation modes above and below $T_C$. The main feature is an increase in the relaxation rates as the temperature is lowered below $T_C$, followed by decrease as the temperature approaches zero. This increase is caused by the appearance of the condensate, and the start of $1 \leftrightarrow 2$ and $1 \leftrightarrow 3$ processes described by ${\cal G}^{12}$ and ${\cal G}^{31}$. The decrease as the temperature approaches zero is caused by the decreasing overall rate factor $\gamma$, which goes as $T^2$. For comparison, we have plotted the relaxation rates of a boson gas above $T_C$, which are obtained from a similar analysis of the Uehling-Uhlenbeck equation \cite{gust2}.

\begin{figure*}[h]
\includegraphics[width=0.9\textwidth]{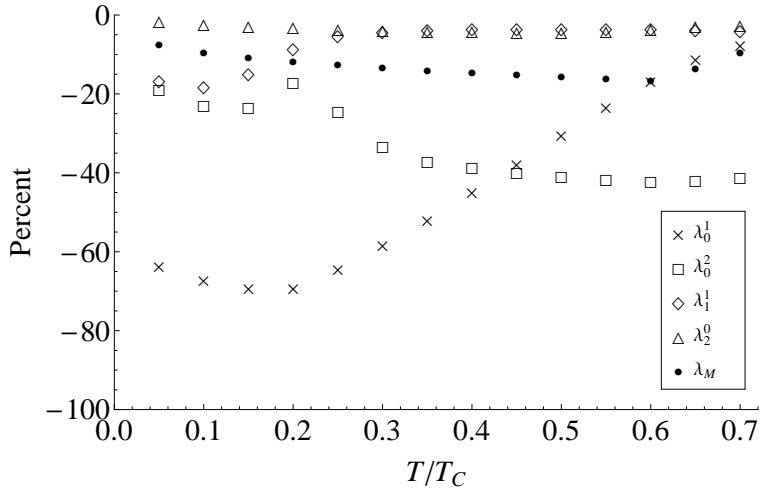}
\caption{Percentage change in several of the lowest non-zero relaxation rates if the effects of ${\cal G}^{31}$ are neglected. Relaxation rates is significantly slower when ${\cal G}^{31}$ is neglected, especially for $\lambda_0^1$, which is the most significant eigenvalue in the relaxation of bogolon number.} \label{fig4}
\end{figure*}

Since the collision integral ${\cal G}^{31}$ is not included in other works, it is of interest to determine its effect on the eigenvalues. To do this, we have computed the eigenvalues without including ${\cal G}^{31}$ and compared them with the eigenvalues when ${\cal G}^{31}$ is included in the calculation. The results of this comparison are seen in Fig. \ref{fig4}, which shows the percent difference in the eigenvalues if the collision integral ${\cal G}^{31}$ is neglected. The neglect of ${\cal G}^{31}$ has a significant effect on the eigenvalues. As expected, neglecting ${\cal G}^{31}$ implies neglecting $1 \leftrightarrow 3$ collisions and results in decreased relaxation rates for all modes.

Of special note is the influence of ${\cal G}^{31}$ on $\lambda_0^1$. To explain this, let us note that above the critical point, there two zero eigenvalues with $l = 0$ and these are associated with particle number and energy conservation. Below the critical point, only one $l = 0$ eigenvalue is zero, which must be associated with energy conservation, since that can be shown explicitly for the collision operator. The smallest non-zero $l = 0$ eigenvalue tends to zero as $T \to T_C$ and is strongly affected by ${\cal G}^{31}$. Since ${\cal G}^{31}$ is does not conserve bogolon number, we interpret the smallest non-zero $l = 0$ eigenvalue as being strongly associated with the relaxation of bogolon number.

\section{Conclusions \label{sec:conc}}

We have considered a new collision integral, ${\cal G}^{31}$, and its influence on the relaxation properties of a gas of bogolons in a dilute Bose-Einstein condensate. To quantify the effects of this collision integral, we have computed the characteristic relaxation rates of the bogolon gas when it is included and when it is neglected. The new collision integral has a significant influence on the relaxation rates and is as important as the two known collision integrals in describing the relaxation of the bogolon gas.

Linearizing the bogolon kinetic equation for small perturbations about equilibrium generates a linear integral equation which can be treated as an eigenvalue equation for the characteristic relaxation eigenmodes and relaxation rates. We have numerically computed the eigenvalues and eigenvectors of the associated matrix equation. These eigenvalues and eigenvectors give us the characteristic relaxation rates of the bogolon gas and a mode expansion for the relaxation of the momentum distribution to equilibrium. We find that the presence of four zero eigenvalues supports the fact that energy and momentum are conserved during during every bogolon collision, while the absence of a fifth zero eigenvalue supports the fact that bogolon number is not conserved.

The characteristic relaxation rates can be related to the transport coefficients. In general, the expressions for the transport coefficients involve a sum over the inverses of the relaxation rates, so no single relaxation rate determines a transport coefficient. From the angular dependence of the eigenfunctions we can make some conclusions based on analogies to the classical Boltzmann Equation \cite{gust1}. Since shear viscosity is related to momentum current correlations, which have a dyadic character, we expect that the $l = 2$ eigenvalues will determine the shear viscosity. Similarly, we expect that the $l = 1$ eigenvalues will determine the thermal conductivity, since it is related to energy current correlations, which have a vector character. We expect that the remaining viscosities of the BEC will depend on the $l = 0$ eigenvalues.

\begin{acknowledgements}
The authors wish to thank the Robert A. Welch Foundation (Grant No. F-1051) for support of this work.
\end{acknowledgements}

\appendix
\section{The Form of the Kernels used in Computation\label{ax:colop}}

To calculate the values of the kernel $K_l(c_1, c_2)$ in Eq. (\ref{eq:Kl}), we split the calculation into six parts, one for each of the individual kernels in Eqs. (\ref{eq:QA} - \ref{eq:QC}). In this appendix, we show how to obtain an expression that is well-suited to numerical quadrature for the kernel $Q_A^l(c_1, c_2)$. Similar procedures can be used with the other five kernels as well. We begin with the definition
\begin{equation}
\begin{split}
Q_A^l(c_1, c_2) =& 2 \pi \int\limits_{-1}^1 d(\hat{\bf c}_1 \cdot \hat{\bf c}_2) P_l(\hat{\bf c}_1 \cdot \hat{\bf c}_2) \int d{\bf c}_3 d{\bf c}_4 \times \\
& \delta^3({\bf c}_1 + {\bf c}_2 - {\bf c}_3 - {\bf c}_4) \delta(\E_1 + \E_2 - \E_3 - \E_4) (W^{22}_{1,2,3,4})^2  \F^0_3 \F^0_4.
\end{split}
\end{equation}
First let us perform the integration over ${\bf c}_4$,
\begin{equation}
Q_A^l(c_1, c_2) = 2 \pi \int\limits_{-1}^1 d(\hat{\bf c}_1 \cdot \hat{\bf c}_2) P_l(\hat{\bf c}_1 \cdot \hat{\bf c}_2) \int d{\bf c}_3 \delta(\E_1 + \E_2 - \E_3 - \E_4) (W^{22}_{1,2,3,4})^2 \F^0_3 \F^0_4.
\end{equation}
The resulting integrand only depends on the magnitude of $c_4$, where $c_4 = |{\bf c}_1 + {\bf c}_2 - {\bf c}_3|$. We now perform the integration over ${\bf c}_3$ in spherical coordinates, with the z-axis oriented parallel to ${\bf c}_1 + {\bf c}_2$,
\begin{equation}
\begin{split}
Q_A^l(c_1, c_2) = 2 \pi &\int\limits_{-1}^1 d(\hat{\bf c}_1 \cdot \hat{\bf c}_2) P_l(\hat{\bf c}_1 \cdot \hat{\bf c}_2)
\int\limits_0^\infty c_3^2 d c_3  \\ \times &\int\limits_{-1}^1 d z_3 \int\limits_0^{2\pi} d \phi_3
\delta(\E_1 + \E_2 - \E_3 - \E_4) (W^{22}_{1,2,3,4})^2  \F^0_3 \F^0_4.
\end{split}
\end{equation}
Now notice that our choice of spherical coordinates for the ${\bf c}_3$ integration allows us to write $c_4$ as
$c_4 = \sqrt{|{\bf c}_1 + {\bf c}_2|^2 + c_3^2 - 2 c_3 |{\bf c}_1 + {\bf c}_2| z_3}$. We can change variables to write the $z_3$ integration as an integration over $c_4$ with $d c_4 = -\frac{2 c_3 |{\bf c}_1 + {\bf c}_2|}{2 c_4} d z_3$,
\begin{equation}
\begin{split}
Q_A^l(c_1, c_2) = 4 \pi^2 &\int\limits_{-1}^1 d(\hat{\bf c}_1 \cdot \hat{\bf c}_2) P_l(\hat{\bf c}_1 \cdot \hat{\bf c}_2)
\int\limits_0^\infty c_3^2 d c_3  \\ \times &\int\limits_{||{\bf c}_1 + {\bf c}_2| - c_3|}^{|{\bf c}_1 + {\bf c}_2| + c_3} d c_4 \frac{c_4}{c_3 |{\bf c}_1 + {\bf c}_2|}
\delta(\E_1 + \E_2 - \E_3 - \E_4) (W^{22}_{1,2,3,4})^2  \F^0_3 \F^0_4.
\end{split}
\end{equation}
Since the integrand now only depends on $c_1$, $c_2$ and $\hat{\bf c}_1 \cdot \hat{\bf c}_2$, let us define $c_A = \sqrt{c_1^2 + c_2^2 + 2 c_1 c_2 (\hat{\bf c}_1 \cdot \hat{\bf c}_2)}$ and use $c_A$ as a change of variables for the $\hat{\bf c}_1 \cdot \hat{\bf c}_2$ integration. This results in
\begin{equation}
\begin{split}
Q_A^l(c_1, c_2) &= 4 \pi^2 \int\limits_{|c_1 - c_2|}^{c_1 + c_2} d c_A \frac{c_A}{c_1 c_2}
P_l\left(\frac{c_A^2 - c_1^2 - c_2^2}{2 c_1 c_2}\right)  \\ \times
&\int\limits_0^\infty c_3^2 d c_3 \int\limits_{|c_A - c_3|}^{c_A + c_3} d c_4 \frac{c_4}{c_3 c_A}
\delta(\E_1 + \E_2 - \E_3 - \E_4) (W^{22}_{1,2,3,4})^2  \F^0_3 \F^0_4.
\end{split}
\end{equation}
To handle the integration over $c_A$, we write the integration limits in terms of Heaviside theta functions,
\begin{equation}
\begin{split}
Q_A^l(c_1, c_2) &= 4 \pi^2 \frac{1}{c_1 c_2} \int\limits_0^\infty c_3 d c_3 \int c_4 d c_4 \int d c_A
P_l\left(\frac{c_A^2 - c_1^2 - c_2^2}{2 c_1 c_2}\right) \delta(\E_1 + \E_2 - \E_3 - \E_4) \\ \times
& (W^{22}_{1,2,3,4})^2  \F^0_3 \F^0_4
\theta(|c_A - c_3| \leq c_4 \leq c_A + c_3)
\theta(|c_1 - c_2| \leq c_A \leq c_1 + c_2)
\end{split}
\end{equation}
and notice that
\begin{equation}
\theta(|c_A - c_3| \leq c_4 \leq c_A + c_3) = \theta(|c_3 - c_4| \leq c_A \leq c_3 + c_4).
\end{equation}
This allows us to move the $c_A$ integration through all of the others and write
\begin{equation}
Q_A^l(c_1, c_2) = 4 \pi^2 \frac{1}{c_1 c_2} \int\limits_0^\infty c_3 d c_3 \int c_4 d c_4 \delta(\E_1 + \E_2 - \E_3 - \E_4) (W^{22}_{1,2,3,4})^2  \F^0_3 \F^0_4 w^l_{1,2,3,4},
\end{equation}
where
\begin{equation} \label{eq:wl}
w^l_{1,2,3,4} = \int\limits_{\max[|c_1 - c_2|, |c_3 - c_4|]}^{\min[c_1 + c_2, c_3 + c_4]} d c_A P_l\left(\frac{c_A^2 - c_1^2 - c_2^2}{2 c_1 c_2}\right).
\end{equation}
The final delta function of energy can now be handled in several ways, but each of them will lead to a well-behaved integrand. Still, we can make a few observations that will help the quadrature go faster.

First, notice that the range of integration on $c_3$ must satisfy $\E_3 < \E_1 + \E_2$. This corresponds to $c_3 < \sqrt{\sqrt{(\E_1 + \E_2)^2 + b^2} - b}$. In fact, the symmetry between $c_3$ and $c_4$ shows that the whole integral is equal to twice the integral from $0 \leq c_3 \leq c_h$ where $c_h = \sqrt{\sqrt{\left(\frac{\E_1 + \E_2}{2}\right)^2 + b^2} - b}$. Also, since it is symmetric in $c_1$ and $c_2$, we can assume that $c_1 > c_2$ in Eq. (\ref{eq:wl}) and swap $c_1$ and $c_2$ if $c_2 > c_1$.

Analysis of the function $w^0_{1,2,3,4}$ shows that under the constraint that $\E_1 + \E_2 = \E_3 + \E_4$, $w^0_{1,2,3,4} = 2 \min[c_1, c_2, c_3, c_4]$. This is only generally true for $Q_A^l$ and not the other kernels. Furthermore, $w^l_{1,2,3,4}$ for $l \geq 1$ can always be written as the product of $w^0_{1,2,3,4}$ and another finite function. The integrand will therefore always have a discontinuity of its derivative when $c_3 = c_2$, and the integration region should be split at this point.

With all of these considerations, we finally write the best form as
\begin{equation}
Q_A^l(c_1, c_2) = \frac{4 \pi^2}{c_1 c_2} \left[ \int\limits_0^{c_2} d c_3 + \int\limits_{c_2}^{c_h} d c_3\right]
\frac{c_3 \E_4}{c_4^2 + b} (W^{22}_{1,2,3,4})^2  \F^0_3 \F^0_4 w^l_{1,2,3,4}
\end{equation}
where $c_4$ takes the value that makes $\E_4 = \E_1 + \E_2 - \E_3$. Three of the other kernels ($Q_B$, $Q_C$, $R_A$) can be similarly reduced to a single integration with a well-behaved integrand, while the other two ($T_A$, $T_B$) can be reduced to explicit functions of $c_1$ and $c_2$.

A final point concerning the kernels involves the function $R_A^0(c_1, c_2)$. Though this function is undefined when $c_1 = c_2$, integrals over the entire kernel $K_l(c_1, c_2)$ still converge. In the numerical method of Sec. (\ref{sec:result}), we use the fact that ${\cal G}^{22}$ acting alone conserves bogolon number to determine the values of $R_A^0(c_1, c_1)$.

\newpage

\end{document}